# OnionNet: a multiple-layer inter-molecular contact based convolutional neural network for protein-ligand binding affinity prediction

*Liangzhen Zheng, Jingrong Fan and Yuguang Mu**

School of Biological Sciences, Nanyang Technological University, 60 Nanyang Drive, Singapore, 637551

*Corresponding author, ygmu@ntu.edu.sg

## Abstract

Computational drug discovery provides an efficient tool helping large scale lead molecules screening. One of the major tasks of lead discovery is identifying molecules with promising binding affinities towards a target, a protein in general. The accuracies of current scoring functions which are used to predict the binding affinity are not satisfactory enough. Thus, machine learning (ML) or deep learning (DL) based methods have been developed recently to improve the scoring functions. In this study, a deep convolutional neural network (CNN) model (called OnionNet) is introduced and the features are based on rotation-free element-pair specific contacts between ligands and protein atoms, and the contacts were further grouped in different distance ranges to cover both the local and non-local interaction information between the ligand and the protein. The prediction power of the model is evaluated and compared with other scoring functions using the comparative assessment of scoring functions (CASF-2013) benchmark and the v2016 core set of PDBbind database. When compared to a previous CNN-based scoring function, our model shows improvements of 0.08 and 0.16 in the correlations (*R*) and standard deviations (*SD*) of regression, respectively, between the predicted binding affinities and the experimental measured binding affinities. The robustness of the model is further explored by predicting the binding affinities of the complexes generated from docking simulations instead of experimentally determined PDB structures.

**1. Introduction**

High binding affinity between a small molecule or a short peptide to a receptor protein is a one of the major selecting criteria in drug discovery [1]. Although the binding affinity could be measured directly through experimental methods, the time cost and finance expenses are extremely high. Therefore, there is an urgent need to develop accurate computational binding affinity prediction models. Several computational methods have been developed to estimate the protein-ligand binding affinity [2,3]. Given the three-dimensional structure of a protein-ligand complex, the binding free energy could be approximated through scoring functions or using Molecular Mechanics Poisson–Boltzmann and surface area continuum solvation (MMPBSA) method and alchemy binding free energy. It is well known that the scoring functions used for binding affinity estimation after docking pose searching are not accurate enough and result in a high false positive rate [4]. While MMPBSA method [5] could provide binding free energies, but not the absolute values, it outperforms the docking scoring functions in general, but it is more time-consuming. Lastly, the alchemy binding free energy estimation [6] is very accurate, however it consumes extremely high computational resources and thus it is not suitable for large scale binding energy estimation during virtual screening.

Generally, the negative logarithms ($pK_a$) of the dissociation constants ($K_d$), half inhibition concentrations (IC50) and inhibition constants ($K_i$) were used to represent the experimental determined binding affinities. Therefore, the performance of "scoring power" was evaluated majorly using two metrics, the Pearson correlation coefficients ($R$) between the experimental $pK_a$ and the predicted $pK_a$, as well as the standard deviations ($SD$) of the regression [7]. The performance of scoring functions has been thoroughly evaluated [7-9]. Based on one of the most popular benchmarks, the comparative assessment of scoring functions v.2013 (CASF-2013, or PDBbind database v2013 core set), the accuracies of the most commonly used scoring functions [7, 8] were compared and evaluated. In addition, the prediction powers of the scoring functions implemented in the two open-source docking packages (AutoDock and AutoDock Vina respectively) [9], were also assessed using CASF-2103 benchmark. Among the scoring functions, X-Score, ChemScore and ChemPLP show the best "scoring power" and "ranking power", while the scoring function implemented in AutoDock Vina shows moderately good "ranking" power. The best scoring function X-score could achieve a $SD$=1.78 and $R$=0.61 with CASF-2013 benchmark [7].

Recent years, another category of predictors, the machine learning (ML) based scoring functions or prediction models emerges as a type of fast and accurate binding affinity prediction methods [10-18]. The early examples such as RF-scores [16] and NNScore [15] generated ML models for binding affinity predictions. RF-score is a random-forest regression model constructed using the inter-molecular interaction features. These two methods had been applied further to re-score the docking results in virtual screening for lead discovery [19, 20].

Different from traditional ML methods, deep convolutional neural networks (CNNs) are more powerful in the sense that they do not rely on experts for feature selections, which is very tricky [21-24]. The non-linear transformations of the raw data set (the three-dimension coordinates of the protein-ligand complex in this case) could uncover the hidden patterns in the data [21, 22]. It thus makes CNNs very suitable not only for image classifications, voice processing and natural language processing, but also for drug discovery [1, 10-12, 21, 25]. CNN models have been applied for assessing whether a specific molecule is a potential binder of a target [26-28]. The performance of such classification models was quite sensitive to the selections of the negative samples (receptor-decoy complexes) [29, 30].

Later, CNN models were adopted for the binding affinity predictions [10-12, 31], and have also been applied for virtual screening [20, 32, 33]. One such model, AtomNet [26], a deep convolutional neural network (CNN) model, took the vectorized grids within a cubic box centered at the ligand as the features for the protein-ligand complex, showing good performance for protein-ligand binding affinity predictions. Other features, such as the protein-ligand topological fingerprints, were also adopted for ML or CNN models [31, 34].

Taking CASF-2013 as benchmark, one of the most accurate binding affinity prediction tools so far is Pafnucy [12], which outperformed other methods in predicting binding affinity, given the three-dimensional protein-ligand complexes structures. For Pafnucy predictor, the chemical information within a box of the size 20 Å× 20 Å×20 Å centered on all ligand atoms was extracted at every 1 Å grid resulting in the 21×21×21×19 high-dimensional data set. Then the dataset was fed to a CNN model [35] and it achieved the best prediction performance ($R$=0.7 and $SD$=1.61) so far.

However, we realize that the interactions collected within the 20×20×20 grid box are rather localized around the ligand. It is well known that the electrostatic interactions, very important in protein-ligand and protein-peptide interactions, are long-range interactions [36, 37] and may not be fully accounted in the cubic box of the size of 20 Å . Meanwhile, the features included in the grid box, such as the atomic partial charges, were calculated using empirical methods like AM1-BCC calculations [38, 39]. These features may not be very accurate and pose noises to the model.

In this study, a different philosophy is applied: non-local protein-ligand interactions are included with minimum bias and noises. To further reduce the orientation biases induced by using features of direct 3D coordinates, the element-specific contacts between proteins and ligands which are internal coordinates and invariant under rotational operations are considered in our model. Such element-specific inter-molecular interaction features in a linear summation form was previously also adopted in the RF-score model [16]. To account for both the local and non-local interactions the contacts between the proteins and the ligands are grouped into different distance ranges. Such protein-ligand interaction features are named as multiple-layer inter-molecular features. We trained a CNN model (named as "OnionNet" hereafter) with the PDBbind v2016 dataset [40] as our benchmark and compared our results with the predictions of different scoring functions (described in the CASF-2013 [7,8]) and Pafnucy [12] using the same standalone CASF-2013 dataset and PDBbind v2016 core set [7]. Our OnionNet model achieves a 1.278 and 1.503 root mean squared error (*RMSE*) for the 290 complexes from PDBbind v2016 core set and 108 complexes from CASF-2013, respectively, outperforms the *RMSE* of 1.42 and 1.69 obtained by Pafnucy. Consistently, the coefficients *R* of 0.812 and 0.786, higher than those of Pafnucy and another model reported by Indra Kundu et al [13], are achieved by our model on these two benchmark datasets.

The robustness of our OnionNet model is tested by inputting predicted protein-ligand complexes structures/poses using docking simulations. The outcoming predicted binding affinities are comparable to those fed with the experimental determined PDB complexes structures. The datasets, the OnionNet model file and necessary pre-processing scripts could be found in the git repository at http://github.com/zhenglz/onionnet/. The codes could be freely modified according to GNU General Public License v3.

**2. Methods and materials**

2.1 Featurization of protein-ligand complexes

The inter-molecular interaction information was extracted from the 3D structures of protein-ligand complexes (Fig. 1). Firstly, we defined a series of boundaries around each atom of the ligand, and the space between boundary $k$-1 and boundary $k$ thus forms a "shell" with a thickness of $\delta$. If $k = 1$, the distance between the atom in the ligand to the nearest point of the boundary is $d_0$, and for boundary $k$ $(when\ 2 \leq k \leq N)$, the minimal distance between the ligand atom to the boundary is $(k-1)\delta+d_0$.

Secondly, the element-pair-specific contact numbers are calculated for the ligand atoms and the protein atoms in each of the $N$ shells. In the original RF-score paper[16], 9 different elements were considered, and one single distance cutoff (1.2 nm) were used, thus it resulted in totally 81 features. The rationale behind this research seems quite straightforward and simple, but RF-score still achieved the state-of-art performance at that time. However, we further considered the possibility to choose different distance cutoffs to include both the short-range and long-range element-specific interactions.

Here in this study, we select 8 elements types ($E_L$), C, N, O, H, P, S, HAX, and Du (Dummy, representing all remaining elements) to quantify the contact types between atoms in ligands and proteins. Here HAX represents any one of the 4 halogen elements F, Cl, Br and I. Although P, HAX and Du may not exist in normal proteins, we keep the elements to maintain the generalization ability of the model. For example, in future, we may incorporate the scoring function to guide the molecular simulations of the ligand binding to the protein with phosphorylation or other types of modifications.

For the shell $n$ (between boundaries $k = n − 1$ and $k = n, 1 \leq n \leq N$), 64 features (considering all possible combinations of elements in a ligand and its target) are used to present the inter-molecular interaction information between the ligand and the protein atoms.

$$E_L = [C, N, O, H, P, S, HAX, DU]$$

$$EC_{T_s,T_t} = \sum_{r=1}^{R_{n,T_s}} \sum_{l=1}^{L_{T_t}} c_{r,l}, \quad \text{while } T_s \in E_L, T_t \in E_L$$

$$c_{r,l} = \begin{cases} 1, & (k-2)\delta + d_0 \leq d_{r,l} < (k-1)\delta + d_0 \\ 0, & d_{r,l} < (k-2)\delta + d_0, d_{r,l} \geq (k-1)\delta + d_0 \end{cases}$$

For any element-pair combination $EC_{T_s,T_t}$, the contact number is the summation of contacts between atom $r$ in shell $k$ of the protein (with element type $T_s$) and atom $l$ in the ligand (with element type $T_t$), while the $R_{n,T_s}$ is the total number of atoms whose element type is $T_s$, and $L_{T_t}$ is the total ligand atom number for type $T_t$. The contact number between atom $r$ and $l$ is 1 if the distance $d_{r,l}$ between them is within the range $(k-2)\delta + d_0 \leq d_{r,l} < (k-1)\delta + d_0$, otherwise 0.

For example, in shell $n$ (between boundary $k$ and $k$-1), the value of the element-pair "C_C" ($EC_{T_s,T_t}, T_s = \text{'C'}, T_t = \text{'C'}$) is the contact number of protein-ligand carbon atom pairs within the distance cut-off ranging from $d = (k-2)\delta + d_0$ and $d = (k-1)\delta + d_0$.

In this study, we define $N$=60 shells with a $d_0$=1.0 Å and $\delta$=0.5 Å. The distance from the farthest boundary ($k$=60) to atoms in the ligand is 30.5 Å. It thus results to 3840 features considering both local and non-local interactions between the protein and the ligand. If converted to a grid box as in Pafnucy [12], the size will be more than 61 Å ×61 Å ×61 Å, 27 times larger than the one used in Pafnucy.

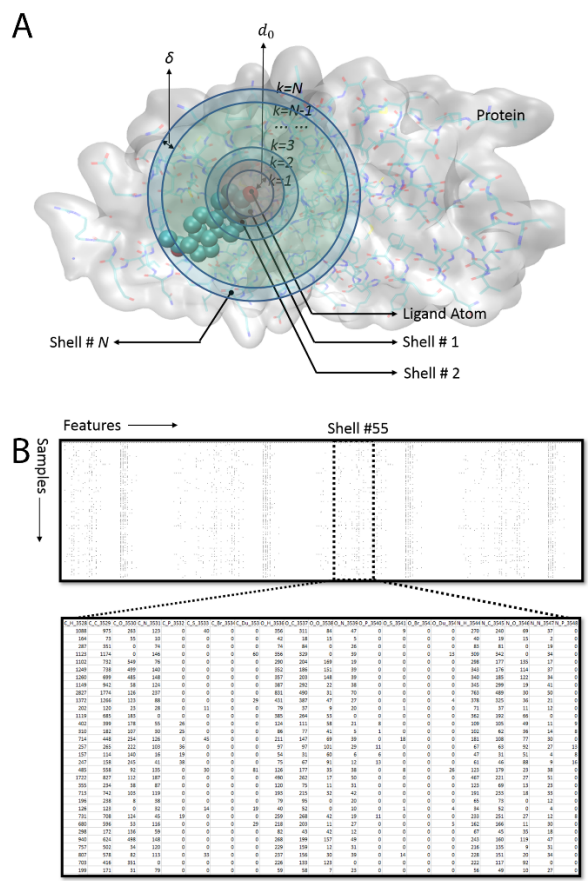

**Figure 1.** Featurization of the protein-ligand complexes based on contact numbers in protein-ligand interaction shells. A) The definition of the "shell-like" partitioning of the protein around the ligand in the three-dimensional space. The PDB ID 1A28 [41] is used as an example here. B) A glimpse of the features of the contact numbers. The features are presented column-wise while the samples are presented row-wise, each row is the information we extracted from one protein-ligand complex, and one column contains a specific feature calculated from all samples.

2.2 Dataset preparation

The OnionNet model was trained and tested with the protein-ligand three-dimensional structures and binding affinities from the updated PDBbind database v2016 (http://www.pdbbind.org.cn/)

(Fig. 2), which was also used by the Pafnucy model. We adopted the same procedure as Pafnucy model. The model was trained with the training set and validating set, while two testing sets were generated for performance evaluation of the model.

There are three overlapping subsets in the original PDBbind v2016 dataset: the core set, the refined set, and the general set. The refined set contains the refined protein-ligand complexes with high-quality binding affinity measurements. The general set contains all the protein-ligand complexes of the PDBbind dataset v2016. Firstly, we extracted all the 290 complexes in the v2016 core set and assigned them into the 1$^{st}$ test set. Then for the remaining complexes in the v2016 refined set, 1000 complexes were randomly selected and used as the validating set. Lastly, the remaining 11906 complexes in the v2016 general set (by removing all complexes in the 1$^{st}$ test set and validating set) were adopted for the training set.

The core set (v2013), or the CASF-2013 benchmark, one subset of the PDBbind database v2013, which was selected by Li et al [7], selects PDB complexes after clustering and is primarily used for validating docking scoring function and CADD benchmark. To compare the performance of our model with other scoring functions conveniently, we prepared 2$^{nd}$ test set containing 108 complexes from v2013 core set by removing the overlapping complexes adopted in the validating and training set. The 2$^{nd}$ test set (called v2013 core set hereafter) are found to be the subset of the v2016 core set (1$^{st}$ test set).

Before protein-ligand complex featurization, we ignored all water molecules and ions. The ligand structures (in mol2 format) were converted to PDB format and combined with the receptor PDB file. To be consistent with previous studies, no further modifications were made to the protein-ligand complexes. A protein-ligand complex structure was first treated by mdtraj [42] and the element types of each atoms thus were determined, and the contact numbers were calculated, as described in the previous section.

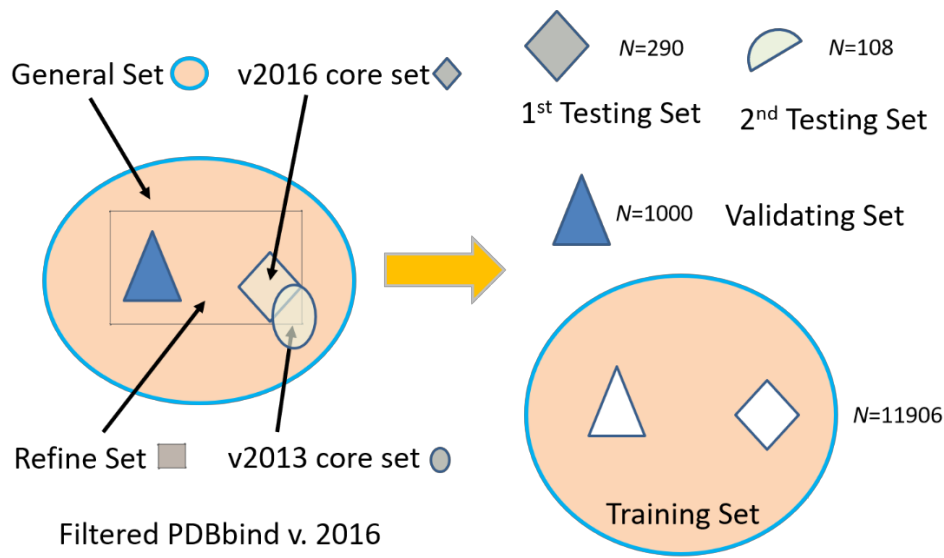

**Figure 2.** The datasets used in the model. The original PDBbind v.2016 dataset was filtered to keep only the protein-ligand complexes, with measured $K_i$ or $K_d$ binding affinities. The remaining filtered dataset thus was divided into 3 disjoint datasets for training, testing and validations. However, two overlapping testing sets were used to compare the performance of our model with other scoring functions. The numbers of protein-ligand complexes have been labeled aside each set in the figure.

To predict the binding affinity, it is a general practice to transform $K_i$ and $K_d$ into the negative log form to train the ML models [12]. In the PDBbind v2016 dataset, the binding affinities of protein-ligand complexes were provided in $K_i$, $K_d$ and IC50. We transformed the binding affinities into $pK_a$ in the following equation (He T. 2017, Simboost):

$$pK_a = -\log_{10} K_x ,$$

where $K_x$ represents IC50, $K_i$ or $K_d$.

Besides using PDB structures, 219 poses with "native-like" structure (RMSD with respect to the native PDB structure less than 2 Å) generated using vina docking software were prepared for model robustness evaluation. The detailed procedures for the docking and pose selections are described in the Support Information (Part 4).

2.3 Deep neural network model

A modified deep convolutional neural network (CNN) was constructed. The architecture of the network is summarized in Fig. 3.

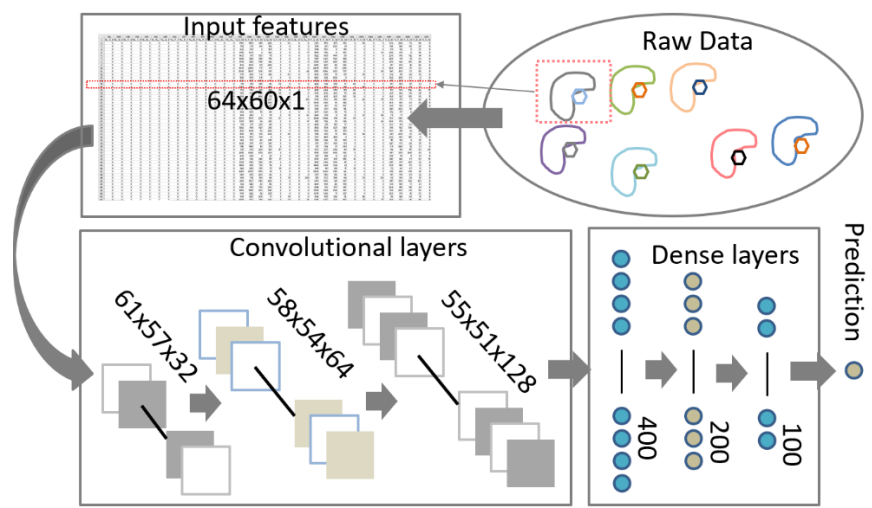

**Figure 3.** The workflow of the protein-ligand binding affinity prediction with OnionNet model.

For each protein-ligand complex, the 3D interactions information is converted into a 2D tensor to mimic a picture with only 1 color channel. The input features thus are fed into the 3-layer convolutional layers, the results thus are flattened and passed to 4 dense layers, and outputs of the last dense layer are transferred to the last layer, the output layer, for $pK_a$ prediction.

The input numerical dataset has 3840 features, which were reshaped to a (64, 60, 1) matrix to mimic an image dataset with only one channel. A sequential model was initialized and followed by 3 two-dimensional convolutional layers. We tried 1D convolutional networks, and the performance is worse than the 2D convolutional models. Regarding the reasons for best performance of the CNN model, we believed that the Y-axis (the different distance-range based shells) has an intrinsic relationship with the X-axis (atom pairs), for example, favorable interactions, such as hydrogen bonds, salt bridges, always require certain atom pair within a certain distance range. CNN may be able to capture such local connections. See more discussions in SI and supplementary table 2.

For the OnionNet model (mCNN-01 in Support Information), there were 32, 64 and 128 filters in the 3 convolutional layers and the kernel sizes were all set as 4, with the strides as 1. No pooling layers attached to the 3 convolutional layers. The results of the last convolutional layer were flattened and passed to the following 4 dense layers with 400, 200 and 100 units. Finally, an output layer was attached with only 1 neuron to generate the predicted $pK_a$. Several different CNN models have been explored (Support Information), and the above-mentioned model achieves the best performance.

A customized loss function was defined to train the model better. During the training of our CNN models, instead of using the default mean squared error (MSE) as the loss function, we designed a new customized loss function to optimize:

$$Loss = \alpha(1 - R) + (1 - \alpha)RMSE,$$

where $R$ and $RMSE$ are the correlation coefficient and root mean squared error respectively, and α is a tunable parameter with positive and less than 1 value. In this study, α=0.8 is used. The rationale is that both high correlation and low root mean squared error are the training target. We found that when α=1.0, the loss function being only determined by $R$, the model has high $R$ value but with high $RMSE$ value as well. The detailed selection of α is described in Support Information.

The kernel sizes were 4, and stride was 1, and no padding was applied in the convolutional layers. For both the convolution layers and dense layers, rectified linear units (ReLU) activation function was adopted [43]. This ReLU function is a fast yet powerful activation function, which has been used in a lot of other deep learning models [44]. ReLU was applied also after the convolutional layers and the dense layers (not including the output layer). Stochastic gradient descent (SGD) optimizer was chosen to search for optimal weights in the model [22, 23]. The learning rate was set as 0.01 with a decay constant $10e^{-6}$ and a 0.9 momentum. Another optimizer, Adam, an extension of the SGD optimizer, was also tested but it made the loss decay very slowly. The batch size (=128) for training was carefully selected (Support information and Supplementary Table 2). Training with small batch sizes renders the model's loss to decay faster but also inducing overfitting issues[45]. Batch normalization was added to each layer except the last output layer [46]. L2 regularization was added to the convolutional layers and dense layers to handle the over-fitting problem. The λ parameter is 0.001, a commonly used value to have a reasonable level of regularization. We screened the optimal dropout probabilities and found that a 0.0 probability in our model achieves the highest prediction accuracy and quick convergence using the validating set, probably because of the usage of batch normalization. Therefore, we did not apply the dropout to the model (with dropout rate = 0.0). Early stopping strategy has been adopted to avoid overfitting issue by holding the training when the validating set loss changes small than 0.01 after a certain number of epochs ($N_{unchange}$=40) (Support Information). The training of the models was based on Keras [47] with Tensorflow [48] as backend.

2.4 Evaluation metrics

Several evaluation metrics were used to assess the model accuracy including the *RMSE*, which quantifies the relative deviations of the predicted values to the experimentally determined values by summing up all squared residuals for each of samples and dividing by number of samples and then computing the square root to have the same physic unit as the original variable (*pK$_a$* in this study).

$$RMSE = \sqrt{\frac{1}{N}\sum_{i=1}^{N}(pK_{a_{predict}} - pK_{a_{true}})2}$$

We also calculated another metrics, standard deviation (*SD*) of the regression, which was also adopted in CASF-2013 benchmark [7] and Pafnucy [12].

$$SD = \sqrt{\frac{1}{N-1}\sum_{i=1}^{N}((a*pK_{a_{predict}} + b) - pK_{a_{true}})^2}$$

where *a* and *b* are the slope and interception of the linear regression line of the predicted and measured *pK$_a$* data points.

Mean absolute error (*MAE*) is another useful evaluation measurement. Different from *RMSE*, *MAE* is the average of the summed absolute differences of the prediction values to the real values.

$$MAE = \frac{1}{N} \sum |pK_{a_{predict}} - pK_{a_{true}}|$$

And finally, the $R$ was another evaluation metrics. It is generally introduced to estimate the correlation relationship between two variables, therefore the predicted $pK_a$ and the real $pK_a$ in this research.

$$R = \frac{E[(pK_{a_{predict}} - \overline{pK_{a_{predict}}})(pK_{a_{true}} - \overline{pK_{a_{true}}})]}{SD_{\overline{pK_{a_{predict}}}} \cdot SD_{\overline{pK_{a_{true}}}}}$$

where $SD_{\overline{pK_{a_{preidct}}}}$ and $SD_{\overline{pK_{a_{true}}}}$ are the standard deviations of the predicted $pK_a$ and the real $pK_a$. The bar notation indicates the mean value of $pK_a$.

## 3. Results

The customized loss, $RSME$ and $R$ were monitored during the training process of the OnionNet model. The best model was obtained with a minimal loss for the validating set at epoch = 89 (Support Information). The prediction accuracy of the model has been accessed based on the following evaluation metrics: $RMSE$, $SD$, $MAE$ and $R$.

Our model achieves correlation coefficients higher than 0.7, and a relatively small $RMSE$ (1.287, 1.278 and 1.503) on the validating set and two testing sets (Table 1). The predicted $pK_a$ and measured $pK_a$ are highly correlated for the two testing sets and validating set (Fig. 4). The accumulated absolute error curves of the validating and testing sets suggest that ~60% and ~50% of the samples have small deviation (~1.0) of $pK_a$ from the measured $pK_a$. The peak of the $\Delta RMSE$ distribution is around 0.4 and 0.7 for the validating and testing sets respectively (Fig. S2 in Support Information).

Table 1. The performance of OnionNet model on different datasets.

| Dataset | R | RMSE | MAE | SD |
|---------|-----|------|------|------|
| Traing set | 0.989 | 0.285 | 0.219 | 0.274 |
| Validating set | 0.781 | 1.287 | 0.983 | 1.282 |
| v2016 core set | 0.816 | 1.278 | 0.984 | 1.257 |
| v2013 core set | 0.782 | 1.503 | 1.208 | 1.445 |

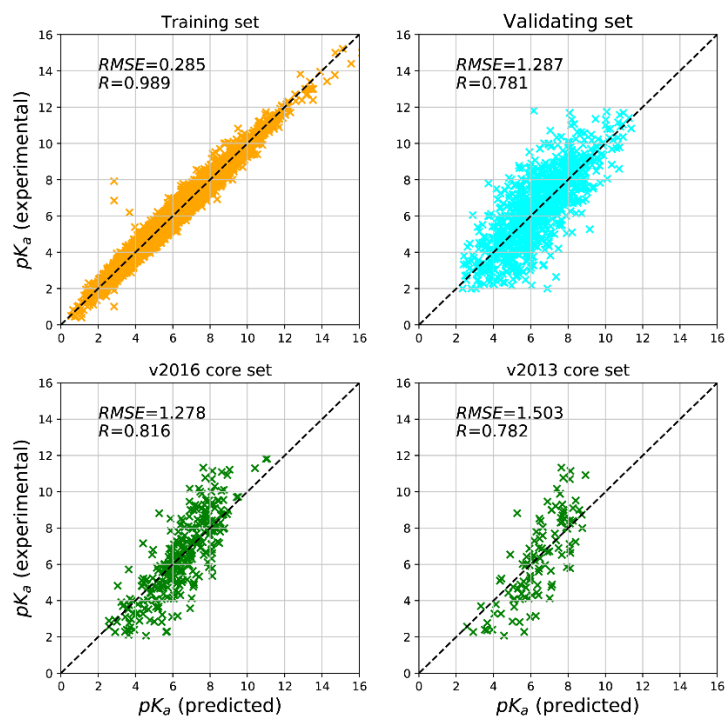

**Figure 4.** The scatter plots of the OnionNet the predicted $pK_a$ against the experimental measurements determined $pK_a$.

## 4. Discussion

4.1 Performance comparison with different scoring functions

Traditional ML models for protein-ligand binding classifications and binding affinity predictions heavily relied on the feature design and selections [49]. The often-adopted protein-ligand binding fingerprints include 3D dimensional raw structural models and/or the amino acid sequences and ligand cheminformatics data, such as the atomic orbitals, hybridization states, atomic charges and molecular topological information [11, 12]. Taking the atomic charges as an example, the hybrid empirical methods, such as AM1-BCC charges, are usually adopted to calculate the "partial charge" of each atom without considering the solvent environment and dipole moments [12]. In this study, we employed simple features without many hypothesis and estimations. The distance-based contacts and chemical element type of each atoms (from both the protein and the ligand) are the only information considered. There are majorly a few advantages to use the distance-based contacts: (1) fewer features would be generated; (2) minimum bias or noise would be introduced; (3) large space around the ligand and both the local and non-local protein-ligand interactions would be taken into consideration; (4) they are internal coordinates and invariant under rotational operations.

The intuitive "simple" features, however, outperform other complicated features-based ML or CNN models (such as OnionNet, Pafnucy, RF-Score, and kNN-Score) [10-13, 17]. Taking the CASF-2013 "scoring power" benchmark as the testing set, the OnionNet model obtained larger $R$ and smaller $RMSE$, $MAE$ and $SD$ than Pafnucy model based on the two testing sets (Table 1). The comparisons between the performance of the OnionNet model and Pafnucy, and other scoring functions are provided in Table 2. The ML and CNN based scoring functions (OnionNet, Pafnucy, RF-Score, and $k$NN-Score) achieve higher accuracies than the popular classic scoring functions (X-Score, ChemScore, ChemPLP, AutoDock Vina score and AutoDock score). The OnionNet model obtained the best correlations between predicted $pK_a$ and the experimentally measured $pK_a$ and got a 0.16 improvement of $SD$ compared with Pafnucy based on the 2nd testing set (v2013 core set).

To demonstrate the statistical reliability, our model has been independently trained for many times. The standard deviations of the $R$ and $RMSE$ of our model are relatively small (Supplementary Table 2). A t-test was performed by comparing the $R$ values of our repeated runs with 0.7 ($R$ value of Pafnucy), assuming the null hypothesis: the average $R$ value of our OnionNet model repeated runs are not higher than $R=0.7$. The one-tail p-value of the t-test is around $9.8*10^{-25}$, meaning the null hypothesis can be rejected confidently. Thus, the reliability of the performance of our OnionNet model is statistically approved.

Table 2. Comparison of prediction power of different scoring functions [7, 9, 12, 14] with CASF-2013 benchmark.

| Scoring functions | SD | R |
|---|---|---|
| OnionNet[a] | **1.45** | **0.78** |
| Pafnucy[b] | 1.61 | 0.70 |
| RF-Score[c] | 1.64 | 0.70 |
| kNN-Score[c] | 1.65 | 0.69 |
| X-Score[d] | 1.78 | 0.61 |
| ChemScore[d] | 1.82 | 0.59 |
| ChemPLP[d] | 1.84 | 0.58 |
| AutoDock Vina[e] | 1.90 | 0.54 |
| AutoDock[e] | 1.91 | 0.54 |

a, 108 complexes; b, 108 complexes; c, 164 complexes; d, 195 complexes; e, 195 complexes.

4.2 Feature importance of the element-type combinations and "shell" location

The understanding of the feature influence on model performance is very important for further model optimization. However, neural networks have a reputation for being used as "black boxes" [50], the importance of the feature is "hidden" and not easy to be dug out. Here, we tackle the problem by removing a set of specific features, retraining the model with the missing features in the original training and validating sets and evaluating the performance loss due to the lack of that set of features (See Support Information). A $\Delta Loss$ is defined as the difference between the loss of a model with missing features and the loss of the best model without missing features. The larger the $\Delta Loss$ is, the higher the loss of the model with the missing features is, the more important those features are.

We first explored the stability of the model upon missing features in a specific layer of shells, as well as the importance the features in this shell. From Fig. 5A, the ligand proximal shells (with smaller shell indices, from 1-15) have relatively higher $\Delta Loss$ than the ligand distal shells (with larger shell indices, from 15-60). The larger $\Delta Loss$ thus suggests that the contributions from the ligand proximal shells are more important than the ligand distal shells. This find is quite consistent with our intuition that local interactions such as van der Waals interactions are important. It is worth to mention that the 1st shell is not the most important for the performance of the model, partially due to the fact that there are very few contacts in the first shell and some close steric crashes between the protein and the ligand will harm the interaction. There is the highest peak around shell 11 (5.5-6 Å), indicating that the medium range interactions contribute to the performance of the model significantly. Interestingly, some distal shells, such as 46, 49 and 53 (23-27 Å), also have large contributions which demonstrates the importance of the non-local interactions.

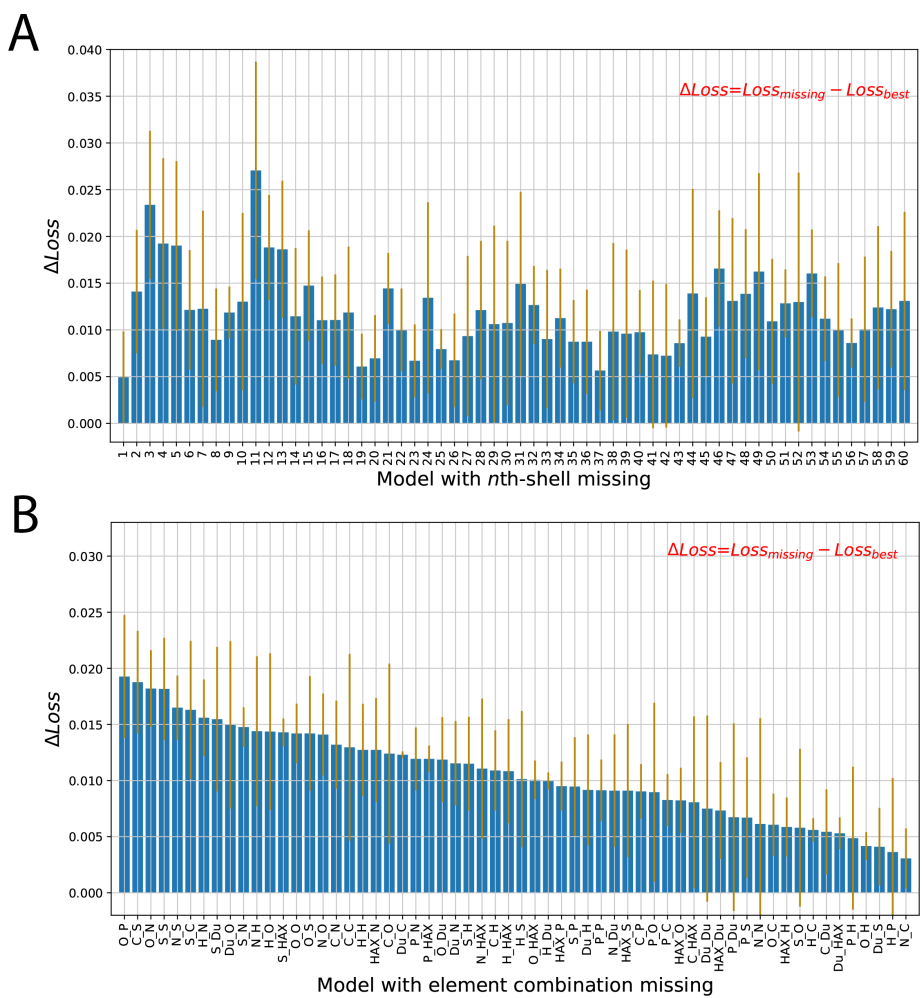

**Figure 5.** The performance change (Δ*Loss*) due to feature missing. A) missing 64 features in a specific "shell" around the ligand. B) missing 60 the same element-type combination in each one of the 60 "shells". The performance change is defined as the difference between the loss of the model with missing features and the loss of the best model. The orange bars indicate the standard deviations of the Δ*Loss* for 5 independent runs.

Furthermore, we explore the feature importance of different element-pair combinations. We iteratively removed 1 type of element-pair combination (out of 64), and then quantified the performance change due to the missing of a set of 60 features (1 feature per shell). The most important element-pair combination is "O_P", which mostly is involved in the strong electrostatic interactions (Fig. 5B). Next, "C_S" is another important element-pair combination, which is involved in the hydrophobic interactions. And the contacts of protein oxygen and sulfur atoms with ligand nitrogen, sulfur or phosphorus atoms also play important roles, while the interactions between protein carbon atoms and ligand hydrogen have minor contributions. The enrichment of sulfur and phosphate related element combinations possibly emphasizes the importance of the less common elements as the "signposts" for input information extraction. On the other hand, the missing of one element-pair combination or one shell of contact interactions does not cause great decreases in the performance of the model which indicates the stability of this model.

4.3 Robustness of the binding affinity prediction model

It is well-known that some classifiers (such as decision tree and deep neural networks) are quite sensitive to the input training data, small changes in the training samples would cause great accuracy lost [51-53]. Thus, there is a risk that training only with the experimental structures may render the model less able to achieve accurate binding affinity prediction when the protein-ligand complex structures were generated from docking simulations, in other words, the robustness of the model may be questionable. To further explore the robustness of the OnionNet model, we directly applied our model to predict the binding affinity of the docking poses for a small set of protein-ligand complexes (Support Information). Docking packages (such as AutoDock Vina) could produce some binding poses with small *RMSDs*. If the *RMSD* between the docking pose and the native conformation is less than 2 Å, the docking pose is called the native-like binding pose. (Fig. 6). The 219 out of 290 docked complexes in PDBbind v2016 core set benchmark were selected for *pK$_a$* prediction, and the *R* and *SD* of the predicted *pK$_a$* against the true *pK$_a$* of this set of complexes are 0.755 and 1.523 respectively (Figure 6A). The performance of the OnionNet model with the inputs from the native-like docking poses are slightly worse than the results obtained directly from the crystal or NMR structures. Taking the pantothenate synthetase (PDB ID: 4DDK) [54] as an example, its ligand 1,3-benzodioxole-5-carboxylic acid (0HN) was extracted from the protein pocket and re-docked back into the pocket using AutoDock Vina and achieved a small *RMSD* = 0.569 Å between the best pose and the native pose of the ligand (Figure 6B). And the binding affinity (*pK$_a$*) of the native pose is 2.29, while with OnionNet, the predicted binding affinity based on the crystal structure and the "native-like" pose is 3.436 and 3.421 respectively. Thus, OnionNet model is found to be robust and insensitive to the small variations of the ligand binding poses.

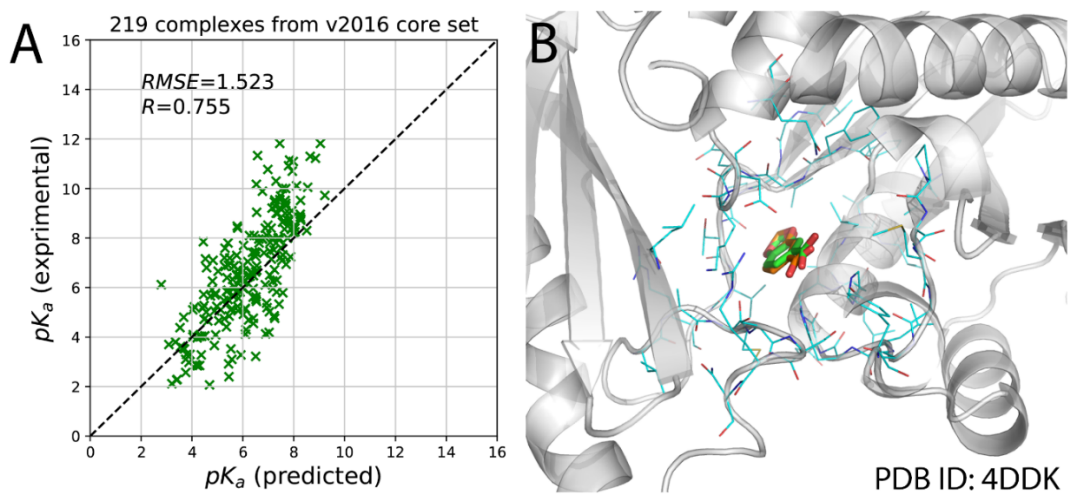

**Figure 6.** The scatter plots of the predicted *pK$_a$* against the experimental determined *pK$_a$* for the selected complexes from v2016 core set and an alignment of a re-docked native-like pose with its native pose. The carbon atoms in the native and native-like "good" poses for the ligand are orange or green respectively, whereas the oxygen and nitrogen atoms are in red and blue. The protein-ligand complex (PDB ID: 4DDK) is one of 219 complexes with native-like poses in the v2016 core set.

**Conclusion**

To accurately predict the binding affinity between the molecule and the target is one of the most important steps in structure-based drug design. To improve ligand binding affinity prediction, we came up with OnionNet model which is based on simple but powerful multiple-layer inter-molecular contact features. The OnionNet model achieves better performance (*R* 0.78 and *RMSE* 1.503) than the current DL-based and classic scoring functions using the CASF-2013 dataset as the benchmark. The stability and robustness of the model were verified through re-training with missing features and predicting the binding affinity on the docking poses. Further improvement of the model would make it a suitable for general lead discovery tasks.

**Support Information**

Support_information.pdf: the file includes detail descriptions as the supplementary information for the method section.

Supplementary Table 1.xls: a spreadsheet containing the PDB codes used for training, validating and testing sets.

Supplementary Table 2.xls: a spreadsheet containing the training and validating performance for multiple trials and hyper-parameter screening.

Supplementary Table 3.xls: a spreadsheet containing the 219 PDB codes with native-like ligand poses generated with AutoDock Vina.

**Corresponding Author**

Yuguang Mu

School of Biological Sciences, Nanyang Technological University, 60 Nanyang Drive, Singapore, 637551

ygmu@ntu.edu.sg

**Author Contributions**

The manuscript was written through contributions of all authors. All authors have given approval to the final version of the manuscript. LZ and YM came up with the plans. LZ and JF performed the experiments and calculations. LZ, JF and YM wrote and revised the manuscripts.

**Funding Sources**

This research was supported by MOE Tier 1 Grant RG146/17.